# Magnetic brightening and control of dark excitons in monolayer WSe$_2$

Xiao-Xiao Zhang[1,2], Ting Cao[3,4], Zhengguang Lu[5,6], Yu-Chuan Lin[7], Fan Zhang[8], Ying Wang[5,6], Zhiqiang Li[5], James C. Hone[8], Joshua A. Robinson[7], Dmitry Smirnov[5], Steven G. Louie[3,4], Tony F. Heinz[2,9]*

**Affiliations:**

[1]Department of Physics, Columbia University, New York 10027, USA

[2]Department of Applied Physics, Stanford University, Stanford, California 94305, USA

[3]Department of Physics, University of California, Berkeley, California 94720, USA

[4]Materials Sciences Division, Lawrence Berkeley National Laboratory, 1 Cyclotron Road, Berkeley, California 94720, USA

[5]National High Magnetic Field Laboratory, Tallahassee, Florida 32312, USA

[6]Department of Physics, Florida State University, Tallahassee, Florida 32310, USA

[7]Department of Materials Science and Engineering and Center for 2-Dimensional and Layered materials, The Pennsylvania State University, University Park, PA 16802, USA

[8]Department of Mechanical Engineering, Columbia University, New York, New York 10027, USA

[9]SLAC National Accelerator Laboratory, 2575 Sand Hill Road, Menlo Park, California 94025, USA

*Corresponding author: tony.heinz@stanford.edu

**Monolayer transition metal dichalcogenide (TMDC) crystals, as direct-gap materials with unusually strong light-matter interaction, have attracted much recent attention. In contrast to the initial understanding, the minima of the conduction band are predicted to be spin split. Because of this splitting and the spin-polarized character of the valence bands, the lowest-lying excitonic states in WX$_2$ (X=S, Se) are expected to be spin-forbidden and optically dark. To date, however, there has been no direct experimental probe of these dark band-edge excitons, which strongly influence the light emission properties of the material. Here we show how an in-plane magnetic field can brighten the dark excitonic states and allow their properties to be revealed experimentally in monolayer WSe$_2$. In particular, precise energy levels for both the neutral and charged dark excitons were obtained and compared with *ab-initio* calculations using the GW-BSE approach. Greatly increased emission and valley lifetimes were observed for the brightened dark states as a result of their spin configuration. These studies directly probe the excitonic spin manifold and provide a new route to tune the optical and valley properties of these prototypical two-dimensional semiconductors.**

The electronic and optical properties of ultrathin TMDC crystals in the MX$_2$ (M = Mo, W, X = S, Se) family have attracted much recent attention. These 2D semiconductors exhibit a direct bandgap at monolayer thickness[1,2], have strong and anomalous excitonic interactions[3-5], and offer the potential for highly efficient light emission. The materials also provide an ideal platform for access to the valley degree of freedom, since the optical selection rules provide a simple means of controlling valley excitation[6-8]. The coupling of the spin and valley degrees of freedom in the valence band (VB) has moreover been recognized as another distinctive feature of the monolayer TMDC systems[7].

Despite important recent advances in our understanding of the electronic and optical properties of these materials, the nature of the spin splitting in conduction bands (CBs) has yet to be fully elucidated by experiments. This information is essential for understanding the radiative properties of the material, since allowed optical transitions in semiconductors occur without change in the spin of the electron. As a result, when the electron spins are polarized along the out-of-plane ($z$) direction, excitons with zero spin ($S_z = 0$) are bright, while excitons with non-zero spin ($S_z = \pm 1$) are dark [9]. In TMDCs, it has been predicted theoretically that the CBs are fully spin polarized, as in the VBs[10-12]. The splitting of the CBs is, however, expected to be relatively modest, with its size comparable to room-temperature thermal energy and, significantly, exhibiting different signs (compared to the valence band splitting) depending on the specific chemical composition of the TMDC crystal. Specifically, in Mo compounds, electrons in the lowest CB are expected to have the same spin as those in the highest VB. In contrast, the opposite spin ordering is expected in W compounds (Fig. 1b), and, correspondingly, the lowest energy exciton is expected to be optically dark. The existence of this lower-lying dark state will quench light emission, particularly at low temperatures. Although distinct thermal activation behavior of light emission has been observed in different TMDC monolayer semiconductors[13-15], in accordance with these predictions, the experiments reveal neither the exact energy structure of the excitonic states nor the conduction band spin arrangement, the latter being of critical importance for valley and spin transport and manipulation in these systems.

Here we mix the spin components of the exciton states in monolayer $WSe_2$ through the application of an in-plane magnetic field. This perturbation relaxes optical selection rules, brightening the dark exciton states and allowing us to observe light emission from these otherwise invisible states directly. This allows us to measure the splitting between the originally bright and dark states with high precision. In the presence of free charge carriers in the $WSe_2$ crystal, trions – charged excitons – also form[16]. With magnetic brightening, the corresponding dark trion states also become visible. Our spectroscopic determination of the energy level structure for the manifold of dark and bright excitonic states agrees well with our predictions from *ab-initio* theory using the GW-BSE formalism. Interestingly, the energy splitting between the bright and dark trion states differs from that for the neutral excitons, reflecting the role of exciton binding energies. Analysis of the splittings provides quantitative information about the CB band structure and many-body interactions in these materials. Brightening the dark excitons also allows us to explore the dynamics and valley properties of these states. Unlike their bright counterparts, we can dramatically alter the radiative lifetime of the dark states by the applied magnetic field and explore states with emission times orders of magnitude longer than for the bright states. Moreover, the measured valley lifetime of the dark excitons, protected from intervalley exchange scattering by their spin configuration, is also significantly larger than the bright exciton valley lifetime, a characteristic of importance for valley pseudospin manipulation [7,17].

Figure 1 provides an overview of the emergence of dark states with an applied in-plane magnetic field. The color plot (Fig. 1a) shows photoluminescence (PL) spectra from an exfoliated $WSe_2$ monolayer for in-plane magnetic fields up to $B_{//} = 31$ T. Two states, labeled as $X_D$ and $X_{DT}$, are seen to emerge and progressively brightened with increasing field. We first present a qualitative picture of how an in-plane magnetic field serves to brighten the spin-forbidden dark exciton states.

The spin-orbit induced splitting of the CB can be attributed to an effective internal magnetic field $B_{int}$, oriented perpendicular to the plane of the 2D layer and acting on the electron spin (Fig. 1b). If we now apply an external *in-plane* magnetic field $B_{//}$, the total effective magnetic field acting on the CB electrons, $B_{eff} = B_{int} + B_{//}$, is tilted slightly away from the normal direction by an angle ~ $B_{//}/B_{int}$ in small $B_{//}$ limit (Fig. 1c). Since the expected spin splitting of CBs of a few tens of meV[10, 11] corresponds to $B_{int}$ of hundreds of Tesla, an appreciable tilt angle is achievable for $B_{//}$ of tens of Tesla. On the other hand, the spin splitting in the VB is ~10 times greater than in the CB, so that the $B_{int}$ is correspondingly larger and the tilting for VB could be neglected. Consequently, $B_{//}$ causes the spin state of electrons in the lower CB to have a finite projection on the zero-field state in the upper VB, and radiative recombination becomes weakly allowed for this otherwise forbidden transition (Fig. 1b). Although the oscillator strength induced by $B_{//}$ remains small, emission from the dark states can still be significant at low temperature due to the large occupation number of the lower-lying dark state. We note that such in-plane magnetic fields have also been used to brighten optically dark excitons in quantum dots[18], carbon nanotubes[19], and quantum wells[20], although the physical mechanisms differ.

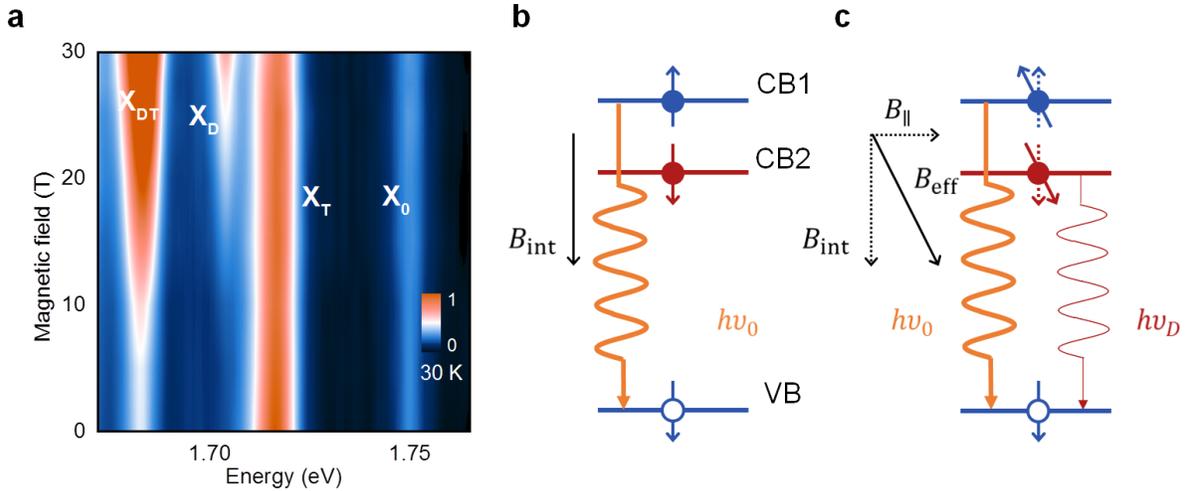

**Fig. 1 | Conduction band structure of monolayer WSe$_2$ and magnetic brightening of dark excitons. a,** False color plot of the measured emission spectrum for monolayer WSe$_2$ at a temperature of 30 K as a function of the strength of the applied $B_{//}$. The displayed energy range includes emission from the neutral A exciton and the associated trion state. Emission from the dark exciton (X$_D$) and dark trion (X$_{DT}$) grows with increasing $B_{//}$. **b,** For monolayer WSe$_2$, electrons in the lower conduction band (CB2) have opposite spin than in the upper valence band (VB), rendering the lowest transition optically dark. Only transitions from the upper CB at energy $h\nu_0$ are allowed. The spin-split CB bands, CB1 and CB2, can be described as the result of an effective out-of-plane magnetic field $B_{int}$ acting on the electron magnetic moment. **c,** Under an external in-plane magnetic field $B_{//}$, the total effective field $B_{eff} = B_{int} + B_{//}$ is tilted away from the surface normal, resulting in tilted spin polarization of the CB electrons. Optical transitions from an excitonic state originating mainly from the lower CB at an energy $h\nu_D$, corresponding to the dark exciton, then become weakly allowed.

Figure 2 shows detailed emission spectra at discrete fields. In the absence of an applied magnetic field, four peaks are visible: the highest-energy peak $X_0$ (1.750 eV) arises from the optically allowed transition of the neutral A exciton; the associated trion state $X_T$ (1.717 eV) is also present due to unintentional doping of the sample[16]; and two lower-energy peaks L1 and L2, attributed to localized excitons associated with defect states[16]. As $B_{//}$ is increased, two additional peaks, labeled as $X_D$ (1.704 eV) and $X_{DT}$ (1.683 eV), emerge and become prominent at high field strengths. On the other hand, the originally bright states show little response to the applied field. As will be justified below in detail, we assign the two peaks $X_D$ and $X_{DT}$ to emission from the neutral dark exciton and the charged dark exciton, respectively.

The WSe$_2$ emission spectrum is simplified at higher sample temperatures where the localized exciton states are absent and the trion feature becomes less prominent, as shown in Fig. 2b for a temperature of 100 K. For $B_{//}$ = 31T, we now see only the emergence of neutral dark exciton $X_D$. The inset of Fig. 2b shows that the extracted $X_D$ feature has the same width as that the $X_0$ peak, which, we note, is significantly smaller than that of the trion. The energy shift between the $X_0$ and $X_D$ features also agrees with that observed in the low temperature data presented above. Additional PL spectra for temperatures between 4 K and 140 K are presented in the Supplementary Information (SI) Section 2.

Additional experimental observations strongly support the assignment of $X_D$ and $X_{DT}$ peaks as neutral and charged dark excitons associated with the split CBs, which are induced by $B_{//}$ according to the scheme described above. First, their emergence depends only on the magnitude of the *in-plane* magnetic field, while an *out-of-plane* field causes only Zeeman energy shifts of the original emission features, with no additional induced emission peaks (as shown in SI Section 3), as reported previously in the literature[21]. Second, the PL intensities of both the $X_D$ and $X_{DT}$ features increase quadratically with $B_{//}$ (Fig. 2c). This is expected: for brightened emission from a spin-forbidden transition, since $B_{//}$ mixes the wavefunction of CB2 with a component from CB1 that is linear in the field strength, it therefore induces a quadratic increase of the dark exciton oscillator strength. We discuss this topic in detail below, as well as the longer emission time observed for the dark trion state compared to the bright state (Fig. 2d). Third, the peak positions are independent of $B_{//}$, consistent with a theoretically predicted shift of < 0.1 meV with the $B_{//}$ up to 31 T (SI Section 7). Fourth, the dark states features are observed in samples of different defect types, and both in exfoliated samples and monolayers grown by chemical vapor deposition (CVD) (SI Section 2). This behavior speaks strongly against the role of defects.

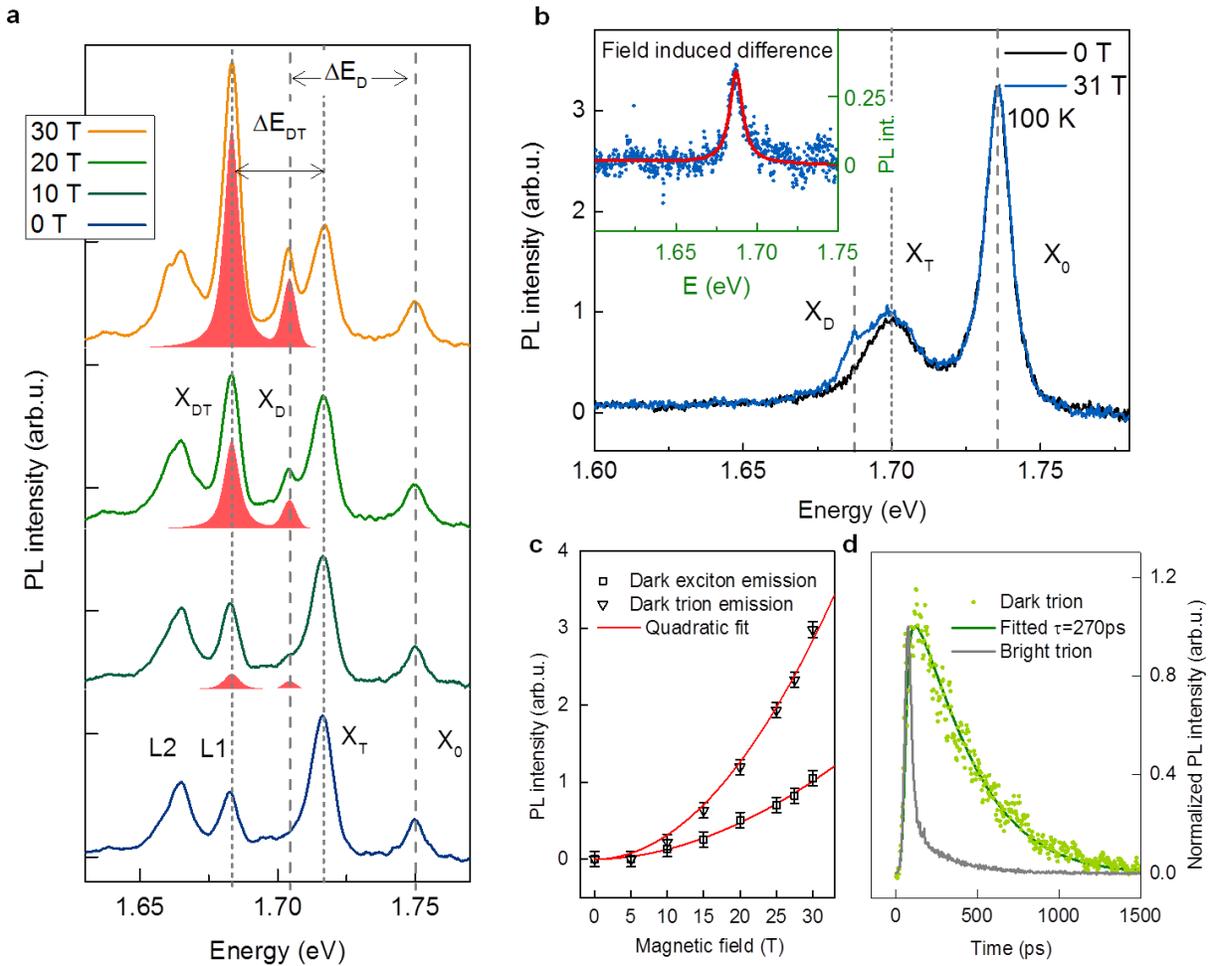

**Fig. 2 | Magnetic field dependence of emission from dark exciton states. a,** PL spectra for selected strengths of the in-plane magnetic field $B_∥$ at a temperature of T = 30 K. For $B_∥ = 0$, the spectrum shows emission from the neutral exciton $X_0$, the trion $X_T$, and localized excitonic states $L_1$ and $L_2$. These features remain unchanged with increasing $B_∥$. Two new peaks grow in at fixed energies, the dark exciton $X_D$ (separated by energy $\Delta E_D$ from $X_0$) and the dark trion $X_{DT}$ (separated by energy $\Delta E_{DT}$ from $X_T$). **b,** PL spectra at T = 100 K for $B_∥ = 0$ and 31 T. For $B_∥ = 0$, the spectrum is dominated by the neutral exciton, with a weaker trion feature. For $B_∥ = 31$ T, the $X_D$ peak emerges, while the $X_{DT}$ emission is too weak to observe. The $X_D$ feature has the same linewidth as $X_0$ (inset), which is about half of the linewidth of the $X_T$ peak. **c,** The PL intensity of the dark exciton and dark trion as a function of $B_{//}$ at T = 30 K. Both features increase quadratically with $B_{//}$. **d,** Comparison of time-resolved PL from the bright and dark trion under excitation by femtosecond laser pulses at T = 4 K, $B_{//} = 17.5$ T. The decay time for the dark state is $\tau = 270$ ps, while for the bright state $\tau < 20$ ps, limited by the instrument response function (44 ps FWHM).

We now discuss the energy shifts of the dark exciton and dark trion relative to the corresponding bright states. Based on the spectra presented above, as well as additional results over a more extended temperature range (4 K to 140 K, see SI Section 2), we find a bright-dark splitting of $\Delta E_D = 47 \pm 1$ meV for the neutral exciton. A similar analysis of the bright-dark splitting for the trion states yields $\Delta E_{DT} = 32 \pm 1$ meV. These values, being comparable to the room temperature thermal energy, are consistent with the observed strong suppression of PL at low temperatures[13-15].

$\Delta E_D$ and $\Delta E_{DT}$ both reflect the splitting $\Delta E_{CB}$, the splitting between the two underlying conduction bands (CB1 and CB 2 in Figs. 1 and 3), but are also affected by the difference in exciton binding energies between the corresponding bright and dark states. Let us first consider how this plays out for the case of the neutral exciton. Using the BerkeleyGW package[22], we applied the *ab-initio* GW method[23] to determine the quasiparticle band structure and used the *ab-initio* GW-BSE approach[24] to calculate the excitonic states of in WSe$_2$ monolayer. Two principal many-body effects contribute to the difference in binding energy: the different spin configuration of the bright and dark excitons, and the different effective masses of electrons in CB1 and CB2. For the former, relative to the $S_z = \pm 1$ dark exciton, the $S_z = 0$ bright exciton experiences an additional repulsive *e-h* exchange interaction, which shifts the bright exciton energy upwards by $E^x$ (Fig. 3a). For the latter, from the quasiparticle band structure, CB2 in WSe$_2$ has a larger mass than higher-lying CB1, as discussed more in details in SI Section 7. This leads to an increased binding energy for the dark exciton; we denote the corresponding mass-induced shift as $\delta E_0$. The bright-dark neutral exciton splitting is thus $\Delta E_D \approx \Delta E_{CB} + E^x + \delta E_0$, the CB splitting combined with the two many-body excitonic corrections. For the overall shift we obtain from the *ab-initio* GW-BSE calculations a value of $\Delta E_D = 57$ meV, in good agreement with the experimental value of 47 meV.

For the bright and dark states of the trion, we need to consider additional many-body effects in this three-body correlated state. For *n*-type trions, which are relevant for our experiment[25], the expected lowest-energy configurations for the bright and dark states are shown in Fig. 3b, as dictated by having opposite electron spins. $\Delta E_{CB}$ still gives the same single-particle contribution to $\Delta E_{DT}$. However, unlike for the neutral excitons, both the bright and dark trions experience *e-h* exchange interactions -- in the bright case, an intravalley exchange ($E^x$) and, in the dark case, an intervalley ($E^{x'}$) exchange interaction (Fig. 3b). As the *e-h* wavefunction overlap of the $S_z = 0$ exciton is almost identical for the intervalley and intravalley configurations, our calculation indicates that $E^x$ and $E^{x'}$ are very similar (within ~ 1 meV), and therefore $\Delta E_{DT}$ can be approximated as $\Delta E_{DT} \approx \Delta E_{CB} + \delta E_T$, where $\delta E_T$ is the mass-induced binding energy difference. Finally, because $\delta E_T$ is expected to be smaller than $\delta E_0$, we expect-- and observe experimentally -- that $\Delta E_D$ exceeds $\Delta E_{DT}$.

From the experimental data and above analysis, we can estimate the CB splitting from our measurements if we assume a mass-induced binding energy difference of $\delta E_T = 0.5\, \delta E_0$, a first-order approximation based on averaging the electron band masses for the trion states. We then have $\Delta E_D - \Delta E_{DT} \approx E^x + \delta E_T$. Taking our experimental data, and noting that $E^x > 0$, we obtain a possible range of conduction band splittings of 17 meV$< \Delta E_{CB} <$ 32 meV. Further, using the theoretical estimate of $\delta E_T \approx 6$ meV, we obtain $\Delta E_{CB} \approx 26$ meV from the relation $\Delta E_{DT} \approx \Delta E_{CB} + \delta E_T$ introduced above. This estimated value of $\Delta E_{CB}$ indicates that electrostatic gating can become effective in modulating the portion of spin for *n*-type carriers within one valley in monolayer WSe$_2$[26].

Up to this point, we have focused on magnetic brightening as a means of observing otherwise dark states and learning thereby about fundamental interactions in monolayer WSe$_2$. Here we demonstrate how this approach can also serve to create new excitonic states with tunable properties. Bright excitons in monolayer TMDC materials exhibit exceptionally short (pico- to sub-picosecond) intrinsic radiative lifetimes, $\tau_{rad-B}(T=0)$, as a consequence of their unusually large binding energy [27]. On the other hand, a long and tunable exciton radiative lifetime would be attractive for many purposes, including precision spectroscopy, control of the valley degree of freedom[6, 17], and the creation of new correlated electronic states[28, 29]. As we now show, brightened dark excitons offer just this possibility. Experimentally, we see evidence for a long radiative lifetime of the brightened dark states in our time-resolved photoluminescence (TRPL) measurements (Fig. 2d). Moreover, the tunability of radiative properties is also directly reflected in the increase in PL strength with $B_\parallel$ (Fig. 2c).

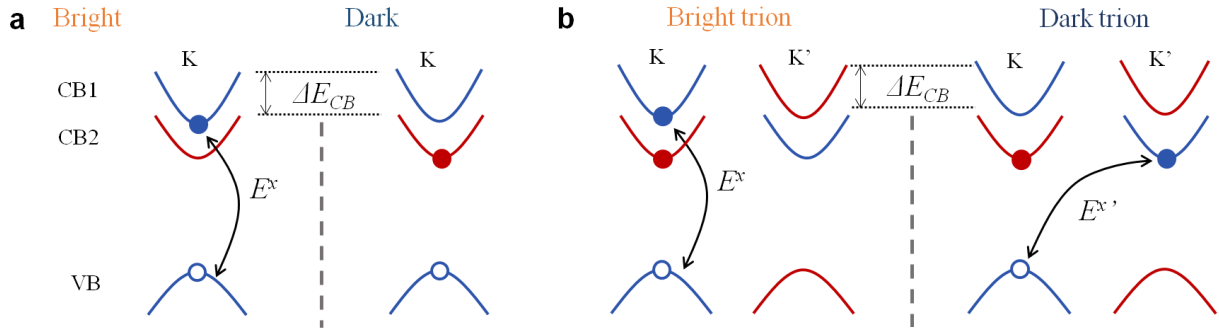

**Fig. 3 | Schematic illustration of bright and dark neutral and charged excitonic states.** The red and blue lines represent spin up and down bands, respectively. **a,** The spin-valley configuration of charge neutral bright and dark excitons. $\Delta E_{CB}$ denotes the energy splitting of the two CBs. The black line represents the *e-h* exchange interactions between the $S_z = 0$ *e-h* pair, which adds to the bright exciton energy by a term $E^x$ that is absent for the dark exciton. **b,** Spin-valley configuration of the lowest energy *n*-type bright and dark trion states. In the dark trion, there also exists an *e-h* exchange interaction term for the intervalley $S_z = 0$ *e-h* pair, as indicated by the black line, with an interaction energy of $E^{x'}$.

Here we describe the effect of the applied $B_\parallel$ on radiative properties of the dark excitons quantitatively. Since the magnetic field-induced mixing occurs predominantly between pairs of dark and bright states, an effective two-level model captures the underlying physics. Influence of the in-plane magnetic field is moreover quite simple because it does not couple to the in-plane motion of electrons, but only to the spin magnetic moment of electrons. In the weak field limit, this model predicts a field-induced oscillator strength for the dark excitons $f_D(B_\parallel) = (\mu_B B_\parallel/\Delta E_D)^2 f_0$, where $f_0$ is the field-free bright exciton oscillator strength, $\Delta E_D$ is the bright-dark state splitting, and $\mu_B$ is the Bohr magneton. (Details of this model are presented in the SI Section 7.) This relation immediately explains the observed quadratic dependence of the PL with $B_\parallel$ (Fig. 2c) under our experimental conditions where non-radiative relaxation of the excitons dominates (and the relaxation channels are not significantly altered by the magnetic field). For the highest applied field of $B_{//} = 31$ T, we predict only a slight brightening of dark excitons of $f_D \approx 1.4 \times 10^{-3} f_0$. As the lower energy state, dark excitons with such small oscillator strength can, however, still dominate in photoluminescence (*e.g.,* Fig.2a) because of their much larger occupation at low temperature. Assuming thermal equilibrium occupation numbers, we can, for example, estimate the dark exciton oscillator strength from the observed (10%) strength of the dark exciton compared to the bright exciton for $B_{//} = 31$ and T = 100 K (Fig. 2b). We infer $f_D \approx 4 \times 10^{-4} f_0$ (see SI Section 7 for details), a value comparable to the prediction of the analysis given above. Moreover, we can estimate the effective radiative lifetime of the brightened dark excitons from the ratio $f_D/f_0$ and the bright-exciton effective radiative lifetime $\tau_{rad-B}(T)$. Taking the previously reported 150 $fs$ as the intrinsic $\tau_{rad-B}$ (T=0), we predict $\tau_{rad-D}$ = 28 ns for T=4 K and $B_{//}$ = 17.5 T corresponding to the conditions of Fig. 2d (SI Section 7). This radiative lifetime is consistent with our measured emission time of 270 ps and the material's low quantum efficiency. For $B_{//}$ = 10 T and T = 50 K, where we can still observe the dark-state emission, we predict $\tau_{rad-D}$ (T = 50 K) $> 1\ \mu s$. For improved crystals exhibiting higher quantum efficiency and narrower emission features, one would be able to observe dark states with still longer radiative lifetimes. We can thus achieve a wide tuning of effective radiative lifetimes under conditions where the brightened dark states are readily observable.

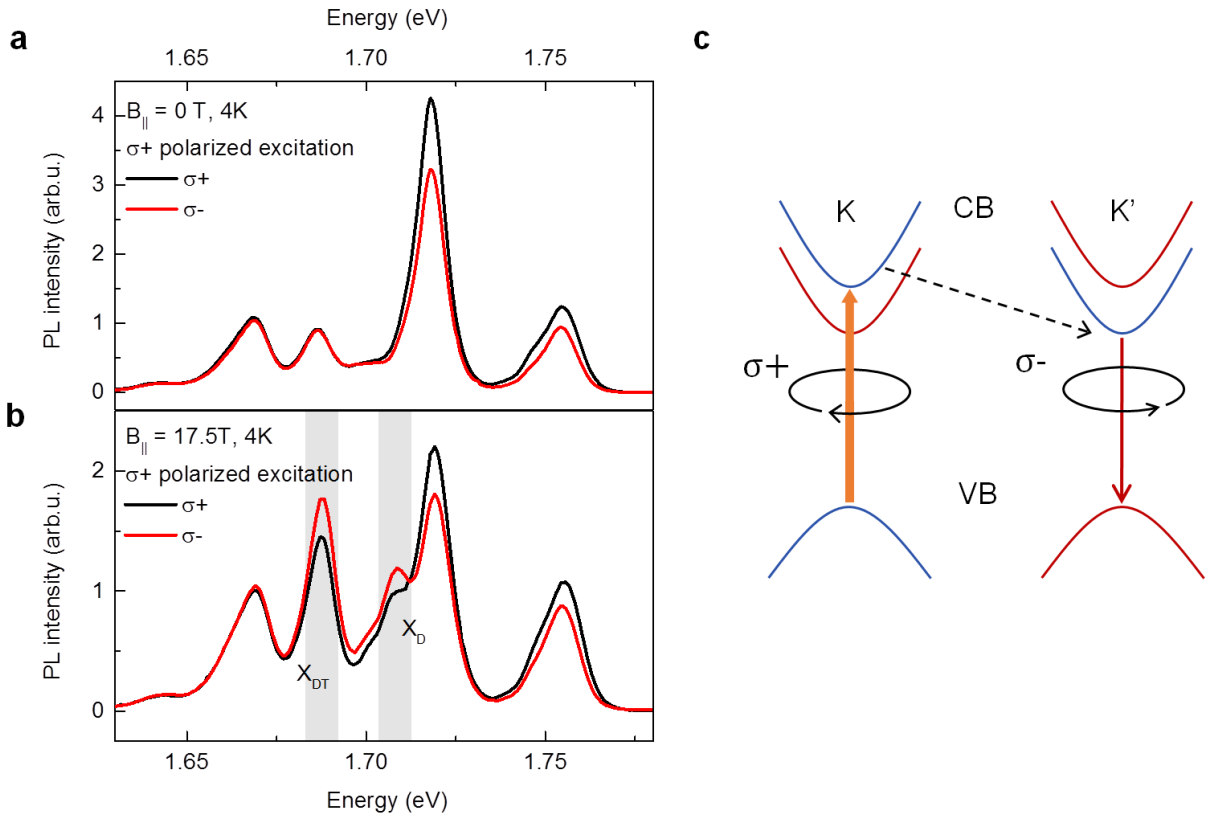

**Fig. 4 | Valley properties of dark excitons based on polarized photoluminescence.** At a sample temperature of 4 K, emission spectra resolved into circularly polarized components are shown for near-resonant excitation (1.88 eV) of the bright exciton with $\sigma^+$ circular polarized radiation for in-plane magnetic fields of **a,** $B_{//}$ = 0 and **b,** $B_{//}$ = 17.5T. For the emission from the bright states, we see enhancement of the emission with the same handedness; for emission from the two brightened dark states (shaded grey areas indicates the energy of $X_D$ and $X_{DT}$ emission), the opposite handedness is increased. **c,** A schematic representation of relaxation through spin-conserving intervalley electron scattering process.

As a final area of exploration, we note that optical access to the valley degree of freedom and the inherent coupling of the valley and spin degrees of freedom in the TMDC monolayers are among the most interesting and distinctive properties of these materials [7]. We have consequently also examined the valley properties of the magnetically brightened dark excitonic states. Since the optical valley selection rules reflect the character of the underlying band states, they should also be obeyed for the magnetically brightened dark states, thus providing a means of probing the valley degree of freedom. Results of such a measurement are presented in Fig 4, using near resonant circularly polarized excitation. For $B_{//} = 0$ (Fig. 4a), the neutral and trion states exhibit enhanced circular polarization in emission matching that of the excitation, and the lower energy defect-related states show negligible circular polarization, as in previous studies[16, 29]. For $B_{//} = 17.5$ T (Fig. 4b), the polarization of the bright states remains essentially unchanged. The new dark exciton $X_D$ and dark trion $X_{DT}$ emission features, on the other hand, show circular polarized emission but with the *opposite* chirality. This intriguing result reflects the distinctive nature of the dark excitons. Although the precise mechanism behind this observation remains unclear, we note, however, the creation of the dark excitons is expected to occur through relaxation of bright excitons by scattering of electrons into the lower CBs, either through an intervalley or intravalley scattering processes. We expect the intervalley process to be more efficient, since it occurs without spin flip. Photoexcitation in one valley, thus gives rise to a higher population of photo-generated electrons in the lowest CB in the other valley, which may explain the observed opposite circularly polarized emission.

The fact that the dark state exhibits circular polarized emission also has implications for the valley lifetime of this state, i.e., the time for intervalley scattering of the dark exciton. Taking the measured dark trion exciton lifetime of ~300 ps (Fig. 2d), together with its measured degree of circular polarization, we infer a dark state valley lifetime exceeding 75 ps. This value exceeds the valley lifetime of the bright excitons[30], in which intervalley exciton exchange between the K and K' valleys is likely to play a crucial role[31, 32]. Scattering of dark excitons between the different valleys, unlike the case of $S_z = 0$ bright excitons, would require a spin flip. Such a process would not occur through an exchange coupling mechanism. The valley lifetime of dark states is therefore protected by their spin configuration.

The approach of magnetic brightening described here has yielded direct spectroscopic information on the energies of spin-forbidden dark excitonic states and spin-orbit band splittings in TMDC monolayers. Importantly, the magnetic brightening of dark excitons also provides a route to produce optically observable states with long and widely tunable radiative lifetimes, as well as greater valley stability compared to the bright excitons. These characteristics are expected to be of considerable value in deepening our understanding of excited states and in the exploration of new quantum states of matter, such as Bose-Einstein condensates, in these materials.

**Methods**

Monolayers of WSe2 were prepared by exfoliation from bulk crystal onto a $SiO_2$/Si substrate with a 300-nm thick oxide layer. The samples were identified by optical contrast microscopy and photoluminescence (PL). The selected monolayer was transferred onto another $SiO_2$/Si substrate with pre-patterned markers to facilitate optical measurements under the magnetic field. Please refer to the SI for grown $WSe_2$ monolayers. Optical measurements in 31T magnet were performed using fiber-based optical probe, while the polarization-resolved PL and TRPL were performed with free-space optics in 17.5T magnet.

Optical measurements in 31T magnet were performed using fiber-based optical probe, while the polarization-resolved PL and TRPL were performed with free-space optics in 17.5T magnet. TRPL measurements were performed by means of the time-resolved single photon counting (TRSPC) technique using the free-space optics arrangement. Excitation was provided by a frequency-doubled mode-locked Ti:sapphire (Coherent, Vitesse) operating at a repetition rate of 80 MHz and providing pulses of 300-fs duration at a wavelength of 400 nm. The collected PL was spectrally filter by a grating monochromator (bandpass of ~10 nm) and detected with a fast avalanche photodiode (PicoQuant, PDM) for analysis using time-resolved single photon counting (PicoQuant, PicoHarp 300). We determined the instrument response function (IRF) of the TR-PL setup by using the temporal profile of produced by the 800-nm femtosecond pulses obtained directly from modelocked Ti:sapphire laser.

Theory: Density functional calculations are performed using the local density approximation (LDA) implemented in the Quantum Espresso package[33]. We use the experimental lattice constant of 3.28 Å through our calculations. The GW[23] and GW+BSE[24] calculations are performed with the BerkeleyGW package[22]. In the calculation of the electron self energy, the dielectric matrix is constructed with a cutoff energy of 35 Ry. The dielectric matrix and the self-energy are calculated on an 18x18x1 k-grid. The quasi-particle band gap is converged to within 0.05 eV. In the calculation of the excitonic states, the quasi-particle band structure is interpolated onto a 180x180x1 k-grid. The electron-hole interaction kernel is also calculated on the same grid. The 1s exciton binding energy is converged to within 0.1 eV. The spin-orbit coupling is included perturbatively within the LDA formalism.